2

# FATORES MACROECONÔMICOS, INDICADORES INDUSTRIAIS E O SPREAD BANCÁRIO NO BRASIL

## *Macroeconomic Factors, Industrial Indexes and Bank Spread in Brazil*

**Carlos Alberto Durigan Junior**
Mestre em Administração pela Faculdade de Economia Administração e Contabilidade (FEA) da Universidade de São Paulo (USP). São Paulo, SP. Brasil. *e-mail: durigancarlos@gmail.com*

**André Taue Saito**
Doutor em Adm. pela Fac. de Econ. Adm. e Contab. (FEA) da Univ. de São Paulo (USP). Prof. Adjunto de Finanças na Escola Paulista de Econ., Política e Negócios (EPPEN) da Univ. Fed. de São Paulo (UNIFESP). São Paulo, SP. Brasil. e-mail: *andretauesaito@gmail.com*

**Daniel Reed Bergmann**
Doutor em Adm. pela Fac. de Econ. Adm. e Contab. (FEA) da Univ. de São Paulo (USP). Prof. do Dep. de Adm. da Faculdade de Econ., Adm. e Contab. (FEA) da Univ. de São Paulo (USP). São Paulo, SP. Brasil.
*e-mail: danielrb@usp.br*

**Nuno Manoel Martins Dias Fouto**
Doutor em Adm. pela Fac. de Econ. Adm. e Contab. (FEA) da Univ. de São Paulo (USP). Prof. do Dep. de Adm. da Faculdade de Econ., Adm. e Contab. (FEA) da Univ. de São Paulo (USP). São Paulo, SP. Brasil.
*e-mail: nfouto@usp.br*

## RESUMO

O objetivo deste trabalho é Identificar os fatores macroeconômicos e os indicadores industriais que influenciaram o spread bancário brasileiro no período de Março de 2011 a Março de 2015. É considerada a subclassificação de alguns segmentos de atividade industrial. Foram utilizados dados mensais de séries temporais em modelos de regressão linear multivariada com uso do Eviews (7.0), dezoito variáveis foram consideradas como possíveis determinantes.. Influenciam positivamente; a inadimplência, os IPIs (Índices de Produção Industrial) de bens de capital, bens intermediários, bens de consumo duráveis, bens semiduráveis e não duráveis, a Selic, o PIB, a taxa de desemprego e o EMBI+. Determinam negativamente; os IPIs bens de consumo e geral, IPCA, o saldo da carteira de crédito e o índice de vendas no varejo. Foi considerado p-valor de 05%.. A conclusão principal é que o progresso da indústria, da geração de empregos e do consumo podem reduzir o spread.

## ABSTRACT

The main objective of this paper is to Identify which macroeconomic factors and industrial indexes influenced the total Brazilian banking spread between March 2011 and March 2015. This paper considers subclassification of industrial activities in Brazil. Monthly time series data were used in multivariate linear regression models using Eviews (7.0). Eighteen variables were considered as candidates to be determinants. Variables which positively influenced bank spread are; Default, IPIs (Industrial Production Indexes) for capital goods, intermediate goods, durable consumer goods, semi-durable and non-durable goods, the Selic, GDP, unemployment rate and EMBI +. Variables which influence negatively are; Consumer and general consumer goods IPIs, IPCA, the balance of the loan portfolio and the retail sales index. A p-value of 05% was considered. The main conclusion of this work is that the progress of industry, job creation and consumption can reduce bank spread.

## 1 INTRODUÇÃO

O *spread* bancário representa a diferença entre as taxas de juros das operações de crédito (taxas de aplicação) e as taxas de captação (BACEN, 2015). O Brasil é apontado como um dos detentores dos maiores *spreads* quando comparado com muitos outros países (JORGENSEN; APOSTOLOU, 2013). O *spread* bancário é internacionalmente utilizado como indicador de eficiência do custo de intermediação financeira (WORLD BANK; IMF, 2005).

Considerando-se a relação do *spread* com o desenvolvimento econômico, as possíveis interfaces entre a atividade industrial e o consumo, a realização deste trabalho emerge da necessidade de identificar se existe influência de fatores macroeconômicos e da atividade produtiva das indústrias brasileiras (esta última medida pelos Indicadores de Produção Industriais IPIs) sobre o *spread* bancário. Esta relação afeta o financiamento e a competitividade da indústria brasileira. Neste contexto, o objetivo principal deste estudo consiste em identificar se o *spread* bancário brasileiro foi influenciado por estes fatores no período de março de 2011 a março de 2015.

O presente trabalho contribui para a literatura, diferenciando-se das demais produções, por explorar a relação da indústria considerando-se a subclassificação industrial (para alguns setores de atividade) e do volume de vendas no varejo (indicador de consumo) com o *spread* bancário no Brasil. Os indicadores das indústrias estão relacionados ao desempenho geral da economia nacional, a demanda por seus diferentes produtos, a demanda por crédito e a expectativas de consumo. Segundo Souza e Coelho (2008) a produção industrial afeta o *spread* bancário, considerando-se que a volatilidade da Selic afeta o nível de produção real e um baixo crescimento pode elevar a inadimplência dos empréstimos (SOUZA; COELHO, 2008).

Considerando esta constatação na literatura, encontramos subsídios para testar a hipótese central do presente trabalho, ou seja, se fatores macroeconômicos e a produção das indústrias afetam o *spread* bancário no Brasil. . Foram utilizados dados do Banco Central do Brasil e o EMBI+ (JP Morgan) como *proxy* para o risco país. A metodologia emprega a regressão linear multivariada com uso do programa Eviews (versão 7.0) e aplica o método Box-Jenkins para testes e análises de possível autocorrelação existente nos resíduos.

O *spread* analisado refere-se ao *spread* total das operações de recursos livres e recursos direcionados (pessoas física e jurídica) do tipo *ex-ante*. Modelos de regressão foram previamente considerados na literatura para estudos sobre *spread* bancário e seus possíveis determinantes (DEMIRGUÇ-KUNT; HUIZINGA, 1999).

Ao todo foram utilizadas dezoito variáveis, são elas; IPCA, IGPM, IPI geral, IPI máquinas agrícolas, IPI extrativa-mineral, Inadimplência total, IPI indústria da transformação, IPI bens de capital, IPI bens intermediários, IPI bens de consumo, IPI bens de consumo duráveis, IPI bens semiduráveis e não duráveis, Selic, PIB, saldo da carteira de crédito total para recursos livres, índice do volume de vendas no varejo total Brasil, taxa de desemprego (região metropolitana) e o EMBI+ (JP Morgan) como variável *proxy* para o risco país.

Os resultados indicam que cinco variáveis determinam negativamente, ou seja, há um mecanismo estatístico que ao ocorrer maior expressão destas variáveis, há redução do *spread* bancário em contexto econômico, são elas; IPCA, o IPI bens de consumo, o IPI geral, o saldo da carteira de crédito total para recursos livres e o índice de volume de vendas no varejo.

Nove o determinam positivamente, ou seja, há um mecanismo estatístico que ao ocorrer maior expressão destas variáveis, há aumento do *spread* bancário em contexto econômico, são elas; a inadimplência total, o IPI para bens de capital, IPI bens intermediários, IPI bens de consumo duráveis, IPI bens semiduráveis e não duráveis, a Selic, o PIB, a taxa de desemprego região metropolitana e o EMBI+. Ao decorrer deste trabalho, os mecanismos serão melhor detalhados e ilustrados.

Foi considerado p-valor de 05% para as variáveis, conforme padrão adotado pela literatura. Foram construídos quatro modelos, os dois primeiros considerando variáveis macroeconômicas e os diferentes indicadores industriais. O segundo modelo, porém, sem o IPGM e com a defasagem da inflação (IPCA) até o terceiro período. O terceiro modelo considera variáveis macroeconômicas, incluindo o PIB. No quarto modelo, além das variáveis macroeconômicas, foi considerado o índice do volume de vendas no varejo como indicador de consumo. Os resultados sugerem que os progressos da indústria, da geração de empregos e do consumo reduzem o *spread* bancário. Tal relação não foi encontrada para os setores de commodities e mineração, estes importantes segmentos da economia brasileira.

## 2 REFERENCIAL TEÓRICO

### 2.1 Conceito

O Banco Central do Brasil (BACEN) define o *spread* bancário como a diferença entre as taxas de juros das operações de crédito e as taxas de captação. O cálculo da composição do *spread* é demonstrado no Relatório de Economia Bancária e Crédito (REBC) disponível no endereço eletrônico do Banco Central do Brasil (BACEN, 2015). O *spread* pode ainda ser utilizado como um indicador de eficiência econômica, quando assume valores muito elevados há indícios de ineficiências no processo de intermediação financeira (WORLD BANK; IMF, 2005).

Em relação à classificação do *spread*, segundo Leal de Souza (2006) o mesmo pode ser classificado segundo três características principais; I- Abrangência da amostra (em função dos bancos e suas operações de crédito), II- Conteúdo, pode haver ou não receitas de tarifas e serviços, e III- Origem da informação, podendo ser *ex-ante* ou *ex-post.* (LEAL DE SOUZA, 2006). Neste estudo é considerado o *spread ex-ante*.

### 2.2 Origem da Informação

*Spread ex-ante*: É calculado com base nas taxas estabelecidas pelos bancos, geralmente é obtido pela diferença entre a taxa de juros de empréstimo e a taxa de juros de captação. As informações bancárias são coletadas e divulgadas pelo Banco Central. O *spread ex-ante* reflete algumas expectativas do banco em relação ao mercado e pode se referir a todas as operações de crédito (em um determinado período analisado) ou apenas a um subconjunto delas. No Brasil o Banco Central utiliza a taxa média de empréstimos das operações de crédito livres (LEAL DE SOUZA, 2006).

A literatura denota que o *spread ex-ante* tende a ser mais sensível do que o *spread ex-post* em relação às mudanças macroeconômicas e percepções de riscos (DEMIRGUÇ-KUNT; HUIZINGA, (1999); AFANASIEFF; LHACER; NAKANE, (2001); NAKANE; COSTA, (2005)). O *spread ex-ante* possibilita capturar os riscos relacionados ao contexto econômico bem como as perspectivas de inadimplência por parte dos tomadores de crédito. Assim, há certa vantagem dos bancos em relação a prováveis prejuízos financeiros oriundos da assimetria informacional, uma vez que os modelos de precificação e risco consideram a probabilidade de inadimplência.

*Spread ex-post*: O spread bancário *ex-post* é calculado por meio do resultado de intermediação financeira realizado pelas instituições bancárias, utilizando receitas e custos efetivos, é calculado com uso de dados contábeis. Entretanto, uma redução observada no spread *ex-post* não necessariamente representa melhor eficiência no processo de intermediação financeira e ainda segundo Demirguç-Kunt e Huizinga (1999) uma redução observada no spread bancário pode ser influenciada por certa redução da inadimplência (DEMIRGUÇ-KUNT; HUIZINGA, 1999).

### 2.3 Contexto Geral

Na literatura há duas abordagens pioneiras sobre o *spread*, a primeira de Klein (1971) que retrata a indústria bancária como um modelo de monopólio, o banco é uma firma (cuja principal atividade é a produção e serviços financeiros) com poder de fixação da taxa de juros, sendo o responsável pela formação de preços no mercado de crédito (KLEIN, 1971).

A segunda abordagem é de Ho e Saunders (1981) a qual retratou o banco como um intermediário entre o tomador e o emprestador de crédito. Primeiramente, os autores determinaram o *spread* puro, este depende da estrutura de mercado e de fatores de risco inerentes à atividade de intermediação. Em outro momento o *spread* puro foi estimado pela regressão contra indicadores dos fatores de risco. A amostra foi composta por informações de 53 bancos norte-americanos com dados contábeis trimestrais no período de 1976 a 1979 (HO; SAUNDERS, 1981).

Em relação à América latina, mesmo após as reformas econômicas, os spreads bancários ainda são elevados. Segundo Chortareas et al. (2012) o nível de intermediação financeira é baixo na América Latina quando comparado aos níveis internacionais (CHORTAREAS et al., 2012).

Em relação ao Brasil, após a evolução recente do crédito, este país atingiu um nível próximo ao encontrado no Chile sobre a relação da porcentagem do crédito ao setor privado sobre o PIB, exibindo um nível de 68.4% enquanto o Chile apresentou relação de 73.2% (valores máximos no período de 2004 a 2013) (PAIM, 2013).

A partir do surgimento da crise de 2008, ocorreu o início da redução da participação privada sobre o crédito total. Em 2012 surgiu uma nova expansão de crédito de origem pública que resultou em observável superioridade da participação de crédito de origem pública sobre a privada em Junho de 2013 (PAIM, 2013).

## 2.4 Brasil

Um dos estudos pioneiros sobre o spread bancário no Brasil é de Aronovich (1994), o qual estudou o comportamento diferencial entre as taxas de captação e empréstimo (spread bancário ex-ante, sem tarifas) (ARONOVICH, 1994). Oreiro et al. (2006) analisaram a influência dos fatores macroeconômicos sobre o spread bancário no Brasil para os anos de 1995 a 2003 com dados mensais. Foi utilizada a técnica do vetor auto regressivo (VAR) na análise do spread ex-ante, sem tarifas e dados do spread médio das operações de crédito com recursos livres pré-fixados disponibilizados pelo Banco Central do Brasil. A medida fora utilizada por Koyama e Nakane (2002a) e Afanasieff et al. (2002) (OREIRO et al, 2006).

Oreiro et al. (2006) adotaram as seguintes variáveis como explicativas; Índice de produção industrial do IBGE, Selic acumulada no mês e anualizada, Inflação medida pelo IPCA do IBGE, volatilidade da Selic (medida pela variância condicional) e alíquota do compulsório sobre depósitos à vista, seguindo as circulares do Banco Central. Os autores aplicaram o uso da volatilidade da Selic como *proxy* para risco. A adoção da produção industrial também já fora realizada por Aronovich (1994), Koyama e Nakane (2002b) e Afanasieff *et al*. (2002). A produção industrial mostrou ter efeito poder de mercado prevalecendo sobre o efeito inadimplência, foram observados aumentos da demanda por crédito e das taxas (OREIRO *et al*., 2006).

Segundo Afanasieff *et al*. (2002) os fatores macroeconômicos são os mais relevantes para a determinação do *spread* bancário no Brasil (AFANASIEFF *et al*., 2002). A tabela 01 a seguir, adaptada de Dantas *et al*. (2011) apresenta os principais estudos sobre os determinantes do *spread* bancário na literatura brasileira com os respectivos resultados de influência e origem da informação (*ex-ante* ou *ex-post*).

**Tabela 01** Principais estudos sobre os determinantes do *Spread* bancário no Brasil

| Método | Autores | Variáveis explicativas e padrão de significância estatística (influência positiva ou negativa sobre o *spread*) |
|---|---|---|
| *Ex-ante* | KOYAMA e NAKANE (2001a e 2001b) | Selic (+); *spread over treasury* (+); impostos indiretos (+); custo administrativo (+); IGP(+); Produto industrial (–); Requerimento de reserva (+). |
| | AFANASIEFF, LHAGER e NAKANE (2001 e 2002) | Custo operacional (+); captação sem custo de juros (+); receita de serviços (+); IGP (+); crescimento do produto industrial (–); Selic (+); volatilidade da Selic (–); banco estrangeiro (–); IGP (–); crescimento do produto industrial (+); *spread over treasury* (+); impostos indiretos (+). |
| | BIGNOTTO e RODRIGUES (2006) | IPCA (–); Selic (+); custo administrativo (+); risco de juros (+); risco de crédito (+); parcela de mercado (–); liquidez (+); receita de serviços (+); compulsório (+); ativo total (+). |
| | OREIRO *et al*. (2006) | Produto industrial (+); Selic (+); volatilidade da Selic (+). |
| *Ex-Post* | GUIMARÃES (2002) | Participação dos bancos estrangeiros (+); caixa e depósitos (+). |

**Fonte:** Adaptado de Dantas *et al.* (2011).

Para Leal de Souza (2006) as metodologias econométricas utilizadas nos trabalhos que abordam o *spread* bancário são relevantes, existindo muitas diferenças entre as especificações dos modelos utilizados, o que pode gerar diferenças nos resultados (LEAL DE SOUZA, 2006). O estudo de Bignotto e Rodrigues (2006) mostra que o IPCA, indicador nacional de inflação, tem padrão de influencia negativo sobre o *spread*, ou seja, em períodos de inflação elevada o *spread* reduz (BIGNOTTO; RODRIGUES, 2006).

Segundo Afanasieff *et al*. (2002) a estabilização do Plano Real teve sucesso no controle da taxa de inflação e em melhor estabilidade macroeconômica, como consequência as taxas de juros foram reduzidas e o crescimento da produção industrial foi observado no país (AFANASIEFF *et al*., 2002).

Segundo a literatura, alguns fatores que influenciam as ineficiências e o elevado custo do *spread* bancário brasileiro são; as forças institucionais e regulatórias, a baixa competição no mercado bancário e as altas taxas de juros (JORGENSEN; APOSTOLOU, 2013). As altas taxas de juros de curto-prazo, fixadas pelo Banco Central, somadas a capacidade dos bancos em cobrar um preço maior do que o custo marginal de produção dos seus serviços (grau de monopólio), também são fatores que explicam as ineficiências do custo de intermediação de crédito no Brasil (OREIRO *et al*,2012).

Koyama e Nakane (2001) apontam que o nível de atividade econômica medido pelo PIB com quatro defasagens, influencia positivamente o *spread*, uma vez que um PIB maior pode refletir em progresso aquisitivo e mais demanda por crédito pela sociedade (KOYAMA; NAKANE, 2001). Matulovic (2015) observou que o produto industrial gerou certo impacto negativo sobre o *spread*. Para o autor o progresso na atividade econômica tende a reduzir a inadimplência (MATULOVIC, 2015). Almeida e Divino (2013) observam que os efeitos positivos vindos do PIB sobre os *spreads* fazem com que estes sejam mais elevados em períodos de crescimento econômico (ALMEIDA; DIVINO, 2013).

Segundo Manhiça e Jorge (2012) a Inflação, o EMBI+ , o desemprego e a taxa de juros mostraram-se significativos a 1% para o primeiro trimestre de 2000 ao terceiro trimestre de 2010.. O desemprego, representando o risco de crédito e também as expectativas de inadimplência, mostrou ser um fator com influência positiva sobre o nível do *spread*, assim o crescimento do desemprego resultará na elevação do *spread* bancário (influência positiva) (MANHIÇA; JORGE, 2012).

A variável EMBI+ é utilizada como aproximação da avaliação de riscos sobre a economia brasileira, permite que o efeito dos riscos de juros e de crédito seja capturado. Segundo Auel e Mendonça (2011) e Pereira Tavares *et al* (2013) o risco país (EMBI+) mostrou-se significativo e positivamente relacionado com o *spread* de crédito, assim ao aumentar o risco país ocorre também maior probabilidade de inadimplência, o que possibilita influenciar a magnitude do *spread* bancário (AUEL; MENDONÇA, 2011; PEREIRA TAVARES *et al*., 2013).

A lógica deste presente trabalho consiste na hipótese central em testar se há relação de influência dos fatores macroeconômicos e da produção industrial sobre o *spread* bancário brasileiro. A hipótese central do presente trabalho é embasada na constatação de Souza e Coelho (2008) de que a produção industrial afeta o *spread*, considerando-se que a volatilidade da Selic afeta o nível de produção real e um baixo crescimento pode elevar a inadimplência dos empréstimos (SOUZA; COELHO, 2008). Entretanto, não foi adotada a volatilidade da Selic no presente estudo, foram considerados; Selic, IPCA, IGPM e EMBI+. Adicionalmente, este trabalho considera o volume de vendas do varejo (como indicador de consumo) e os IPIs de diferentes segmentos industriais nacionais, contribuindo para a literatura.

## 3 PROCEDIMENTOS METODOLÓGICOS

### 3.1 Métodos

Com o objetivo de identificar os fatores influentes sobre o *spread* bancário foi utilizada regressão linear multivariada. Modelos de regressão foram previamente considerados na literatura para estudos sobre *spread* bancário e seus possíveis determinantes (DEMIRGUÇ-KUNT; HUIZINGA, 1999). Para verificar existência de autocorrelação existente entre as séries utilizou-se o método Box-Jenkins (1970), estes autores consideraram um grupo de processos estocásticos importantes como candidatos a terem gerado a série temporal de interesse. São eles o processo autoregressivo (AR), analisado por Yule em 1927, o processo de médias móveis (MA), estudado por Yule em 1926 e o processo integrado (I).

É possível que uma série temporal tenha sido gerada por mais de um desses processos, conhecidos como modelos ARIMA (autoregressivo integrados e de médias móveis), estes são modelos que conseguem descrever os processos de geração de uma variedade de séries temporais para os previsores (que correspondem aos filtros) não existindo a obrigatoriedade de considerar as relações econômicas que formaram as séries. Segundo Mueller (1996) o modelo de Box-Jenkins pode ser utilizado para séries temporais de quaisquer naturezas (MUELLER, 1996).

## 3.2 Dados

Foram utilizadas séries de dados temporais para análise dos determinantes do *spread* bancário total (recursos livres e direcionados de pessoas físicas e jurídicas) do tipo *ex-ante*. O período de análise compreende dados mensais de Março de 2011 a Março de 2015, a escolha deste período é justificada por mudanças metodológicas relacionadas ao cálculo do *spread* bancário pelo Banco Central do Brasil. Todas as séries vigentes de *spread* bancário em indicadores de crédito na base de séries temporais do Bacen têm início a partir de primeiro de Março de 2011.

Os dados foram majoritariamente obtidos do Banco Central do Brasil (Bacen), por meio de acesso ao Sistema Gerenciador de Séries Temporais (SGS) Versão 2.1 Modelo Público, disponível na plataforma do *website* do Bacen (www.bcb.gov.br). Somente o EMBI+ (calculado pelo JP Morgan), utilizado como *proxy* para o risco Brasil, foi obtido do Ipeadata (www.ipeadata.gov.br). A justificativa da utilização das séries utilizadas neste estudo está embasada na importância macroeconômica e em suas relações com o crédito bancário, como também aponta a literatura utilizada. A tabela 02 a seguir mostra todas as séries utilizadas neste estudo, obtidas do Bacen, para avaliar a capacidade de influência sobre o *spread* bancário.

**Tabela 02** Variáveis independentes de fonte do Bacen

| Variável independente | Nº da série no SGS |
|---|---|
| IPCA | 433 |
| IGP-M | 189 |
| Índice de produção Industrial (IPI) – Geral | 21859 |
| IPI – Máquinas Agrícolas | 1388 |
| Inadimplência total | 21082 |
| IPI – Indústria Extrativa Mineral | 21861 |
| IPI – Indústria da Transformação | 21862 |
| IPI – Bens de Capital | 21863 |
| IPI – Bens intermediários | 21864 |
| IPI – Bens de consumo | 21865 |
| IPI – Bens de consumo duráveis | 21866 |
| IPI – Bens semiduráveis e não duráveis | 21867 |
| Taxa Selic anualizada | 4189 |
| PIB mensal | 4380 |
| Saldo da cart. de créd. total – recursos livres | 20542 |
| Índice de vol. de vendas no varejo tot. Brasil | 1455 |
| Taxa de desemp. região metropolitana Brasil | 10777 |

**Fonte:** Elaborada pelos autores.

O EMBI+ (JP Morgan) utilizado como variável *proxy* para o risco Brasil, foi obtido pelo website do Ipeadata. A variável dependente no estudo foi o *spread* médio total (recursos livres e direcionados para pessoas físicas e jurídicas) no período de março de 2011 a março de 2015, sendo utilizada a série 20783 do Banco Central do Brasil.

## 3.3 Procedimentos

Para o cálculo da regressão linear foi utilizado o programa *Eviews* versão 7.0. As variáveis Selic, inadimplência e *spread* foram modelas em primeira diferença, ou seja, são integradas em 1ª ordem, indicando que a série possui raiz unitária, não sendo estacionária. Séries não estacionárias não podem ser modeladas. Tais variáveis foram apontadas por correlogramas lentamente declinantes. Testes de resíduos foram realizados para observar a possibilidade de existência de auto-correlação existente. Quando há autocorrelação nos resíduos a validade estatística do modelo é negativamente afetada. São apresentados quatro modelos como estatisticamente adequados para a explicação do *spread* bancário. A tabela 03 detalha os componentes dos modelos de regressão.

**Tabela 03** Modelos e Variáveis independentes

| Modelo | Variáveis Independentes |
|---|---|
| 01 | EMBI+, IGPM, Inadimplência total, IPCA, IPI bens de capital, IPI bens de consumo, IPI bens duráveis, IPI bens intermediários, IPI bens semiduráveis e não duráveis, IPI indústria extrativa mineral, IPI geral, IPI indústria da transformação, Índice de produção de máquinas agrícolas e a Selic. |
| 02 | EMBI+, Inadimplência total, IPCA (t-3), IPI bens de capital, IPI bens de consumo, IPI bens duráveis, IPI bens intermediários, IPI bens semiduráveis e não duráveis, IPI indústria extrativa mineral, IPI geral, IPI indústria da transformação, Índice de produção de máquinas agrícolas e a Selic. |
| 03 | Inadimplência total, crédito total, PIB (-4), desemprego e Selic (–1). |
| 04 | Inadimplência total, crédito total, índice de vendas no varejo (–5), desemprego e Selic (–1). |

**Fonte:** Elaborada pelos autores.

De acordo com a tabela acima, os modelos possuem as especificações traduzidas pelas seguintes equações:

**Modelo 01:**

$$d(S)_t = \beta_0 + \beta_1 \text{ embi} + \beta_2 \text{ IGPM} + \beta_3 \text{ d}(inadtotal_t) + \beta_4 \text{ IPCA} + \beta_5 \text{ IPI}_{BC} + \beta_6 \text{ IPI}_{C} + \beta_7 \text{ IPI}_{CD} + \beta_8 \text{ IPI}_{INT} + \beta_9 \text{ IPI}_{NSD} + \beta_{10} \text{ IPI}_{EM} + \beta_{11} \text{ IPI} + \beta_{12} \text{ IPI}_{T} + \beta_{13} \text{ IPMA} + \beta_{14} \text{ Selic} + \varepsilon_t$$

**Modelo 02:**

$$d(S)_t = \beta_0 + \beta_1 \text{ embi} + \beta_2 \text{ d}(inadtotal_t) + \beta_3 \text{ IPCA}(t-3) + \beta 4 \text{ IPI}_{BC} + \beta_5 \text{ IPI}_{C} + \beta_6 \text{ IPI}_{CD} + \beta_7 \text{ IPI}_{INT} + \beta_8 \text{ IPI}_{NSD} + \beta_9 \text{ IPI}_{EM} + \beta_{10} \text{ IPI} + \beta_{11} \text{ IPI}_{T} + \beta_{12} \text{ IPMA} + \beta_{13} \text{ Selic} + \varepsilon_t$$

**Modelo 03:**

$$d(S)_t = \beta_0 + \beta_1 \text{ d}(inadtotal_t) + \beta_2 \text{ d}(creditototal_t) + \beta_3 \text{ d}(PIB_{t-4}) + \beta_5 \text{ d}(desemprego_t) + \beta_6 \text{ d}(selic_{t-1}) + \varepsilon_t$$

**Modelo 04:**

$$d(S)_t = \beta_0 + \beta_1 \text{ d}(inadtotal_t) + \beta_2 \text{ d}(creditototal_t) + \beta_3 \text{ d}(venda_varejo_{t-5}) + \beta_5 \text{ d}(desemprego_t) + \beta_6 \text{ d}(selic_{t-1}) + \varepsilon_t$$

Os modelos 01 e 02 consideram os indicadores das atividades industriais, além das variáveis macroeconômicas. Os modelos 03 e 04 consideram as variáveis: PIB (modelo 03), índice de volume de vendas no varejo (modelo 04), taxa de desemprego para a região metropolitana, saldo de crédito total, inadimplência e Selic. A escolha de tais variáveis é pela importância macroeconômica e relação com o crédito bancário, o PIB como um indicador de produção e vendas no varejo como indicador de consumo.

Crédito total, PIB e desemprego foram modelados em 1ª diferença. Em relação ao critério de significância (p-valor), as variáveis independentes que foram consideradas como influentes sobre o spread bancário são as que apresentam p-valor de até 05%. Embora a maioria das variáveis estatisticamente significantes são significativas a 01%, foi adotado o critério de p-valor a 05%, conforme utilizado na literatura.

## 4 RESULTADOS

No modelo 01 as variáveis estatisticamente significantes e explicativas do *spread* bancário são as que apresentaram p-valor de até 05%. São elas: EMBI+, inadimplência total, IPI bens de capital, IPI bens de consumo (este com padrão e influência negativa) , IPI bens de consumo duráveis, IPI bens intermediários, IPI bens semiduráveis e não duráveis. O teste de resíduo para este modelo mostra que o mesmo é estatisticamente bem especificado, os valores estão dentro do intervalo de confiança. O presente modelo possui um bom valor de $R^2$ (68,19%) e há significância estatística de que certos indicadores de atividade industrial influenciam o *spread* bancário. A tabela 04 e a imagem 01 a seguir, apresentam os detalhes estatísticos do modelo 01.

**Tabela 04** Resultados do modelo 01

| Variável | Coeficiente | Erro padr. | t-Estatística | Prob. |
|---|---|---|---|---|
| C | 1.994050 | 1.212845 | 1.644110 | 0.1096 |
| EMBI | 0.004152 | 0.001435 | 2.893976 | 0.0067 |
| IGPM | –0.043672 | 0.122942 | –0.355226 | 0.7247 |
| D(INADTOTAL) | 3.581371 | 0.778621 | 4.599633 | 0.0001 |
| IPCA | 0.078989 | 0.282952 | 0.279160 | 0.7819 |
| IPI_BENSCAPITAL | 0.150154 | 0.054025 | 2.779352 | 0.0089 |
| IPI_BENSCONSUMO | –3.668246 | 1.466077 | –2.502084 | 0.0175 |
| IPI_BENSCONSUMODUR | 0.968882 | 0.347294 | 2.789800 | 0.0087 |
| IPI_BENSINTERMEDIARIOS | 0.942314 | 0.339171 | 2.778283 | 0.0089 |
| IPI_BENSONAOSEMIDUR | 3.202170 | 1.114755 | 2.872533 | 0.0071 |
| IPI_EXTRATIVA_MINERAL | –0.070110 | 0.062574 | –1.120445 | 0.2706 |
| IPI_GERAL | –1.244384 | 0.709924 | –1.752842 | 0.0889 |
| IPI_TRANSFORMACAO | –0.311361 | 0.504486 | –0.617185 | 0.5413 |
| PROD_MAQUINASAGRIC | 4.05E-06 | 5.77E-06 | 0.701706 | 0.4878 |
| D(SELIC) | 0.265726 | 0.208545 | 1.274187 | 0.2115 |
| | | | | |
| R-quadrado | 0.681947 | Média dependente var. | | –0.033333 |
| R-quadrado ajustado | 0.547015 | S.D. dependente var. | | 0.481281 |
| S.E. regressão | 0.323922 | Akaike info. critério | | 0.833677 |
| Sum quadrados res. | 3.462536 | Schwarz critério | | 1.418428 |
| Log probabilidade | –5.008259 | Hannan-Quinn critério. | | 1.054655 |
| F-estatística | 5.054022 | Durbin-Watson stat. | | 2.061200 |
| Prob(F-estatística) | 0.000064 | | | |

**Fonte:** Os autores (resultados gerados pelo *Eviews*)

**Imagem 01** Correlograma dos resíduos do modelo 01

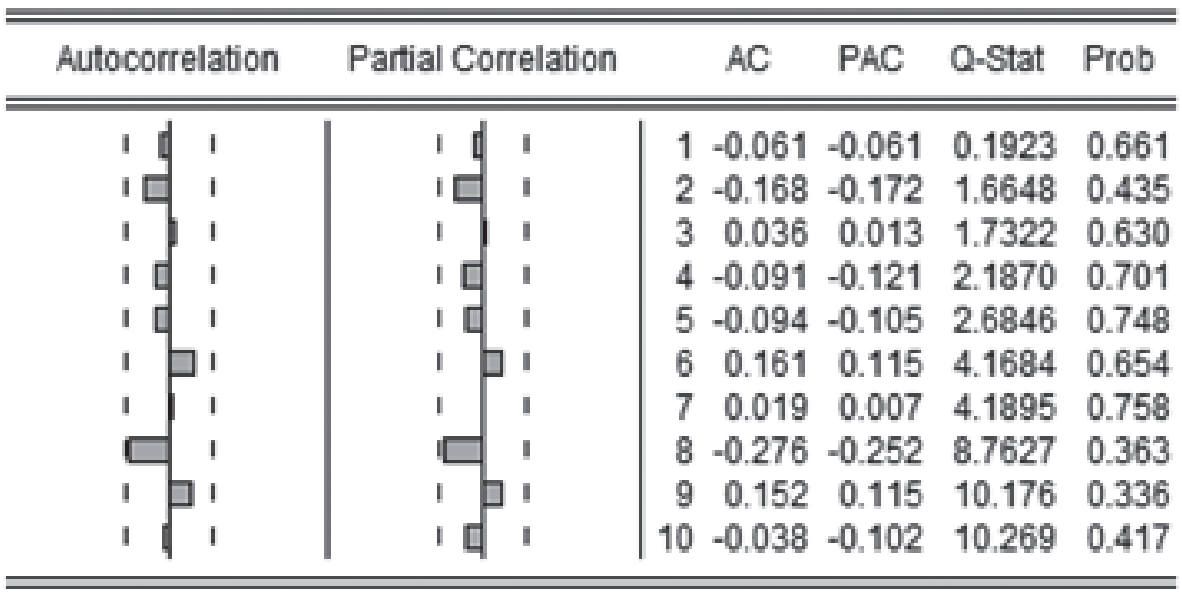

| Autocorrelation | Partial Correlation | | AC | PAC | Q-Stat | Prob |
|---|---|---|---|---|---|---|
| | | 1 | -0.061 | -0.061 | 0.1923 | 0.661 |
| | | 2 | -0.168 | -0.172 | 1.6648 | 0.435 |
| | | 3 | 0.036 | 0.013 | 1.7322 | 0.630 |
| | | 4 | -0.091 | -0.121 | 2.1870 | 0.701 |
| | | 5 | -0.094 | -0.105 | 2.6846 | 0.748 |
| | | 6 | 0.161 | 0.115 | 4.1684 | 0.654 |
| | | 7 | 0.019 | 0.007 | 4.1895 | 0.758 |
| | | 8 | -0.276 | -0.252 | 8.7627 | 0.363 |
| | | 9 | 0.152 | 0.115 | 10.176 | 0.336 |
| | | 10 | -0.038 | -0.102 | 10.269 | 0.417 |

**Fonte:** Os autores (resultados gerados pelo Eviews)

Em relação ao presente modelo pode ser observado que a inflação (medida nacionalmente pelo IPCA) e a taxa básica de juros (Selic) não foram estatisticamente significantes. Considerando isto, foi aplicado o modelo 02 (baseado no modelo 01, excluindo o IGPM, pois pode haver correlação entre este e o IPCA) com defasagem para a inflação, a qual é estatisticamente significante em três defasagens (p- valor a 05%) e com sinal negativo, ou seja, um aumento do IPCA implica em redução do *spread* após três meses.

Na prática esta observação parece ser economicamente contraditória, o que pode indicar que a inflação não afeta o *spread* diretamente, entretanto há um sentido econômico lógico entre inflação e inadimplência, em períodos de alta inflação pode ocorrer aumento da inadimplência e consequentemente elevação do *spread*.

A inadimplência é significante e positiva no modelo 01. O IPI geral é estatisticamente significante e com padrão negativo de influência neste segundo modelo. A Selic também é estatisticamente significante quando o IPCA é defasado em três períodos, apresentando influência positiva sobre o *spread*, ou seja, a Selic elevada tende a influenciar o aumento do *spread*. A tabela 05 e a imagem 02 a seguir, apresentam os detalhes estatísticos do modelo 02 ($R^2$ de 72,09%).

**Tabela 05** Resultados do modelo 02

| Variável | Coeficiente | Erro padr. | t-Estatística | Prob. |
|---|---|---|---|---|
| C | 3.019877 | 1.072811 | 2.814921 | 0.0083 |
| EMBI | 0.003114 | 0.001466 | 2.123488 | 0.0415 |
| D(INADTOTAL) | 3.847871 | 0.732662 | 5.251905 | 0.0000 |
| IPCA (t-3) | –0.603743 | 0.293108 | –2.059797 | 0.0476 |
| IPI_BENSCAPITAL | 0.172377 | 0.048394 | 3.561941 | 0.0012 |
| IPI_BENSCONSUMO | –3.623745 | 1.375933 | –2.633663 | 0.0129 |
| IPI_BENSCONSUMODUR | 0.968326 | 0.326482 | 2.965943 | 0.0057 |
| IPI_BENSINTERMEDIARIOS | 1.063219 | 0.305581 | 3.479338 | 0.0015 |
| IPI_BENSONAOSEMIDUR | 3.204313 | 1.047870 | 3.057929 | 0.0045 |
| IPI_GERAL | –1.362380 | 0.671810 | –2.027925 | 0.0510 |
| IPI_TRANSFORMACAO | –0.383612 | 0.506012 | –0.758109 | 0.4539 |
| PROD_MAQUINASAGRIC | 3.51E–06 | 5.64E–06 | 0.623221 | 0.5376 |
| IPI_EXTRATIVA_MINERAL | –0.074119 | 0.062501 | –1.185874 | 0.2444 |
| D(SELIC) | 0.426728 | 0.212428 | 2.008811 | 0.0531 |
| | | | | |
| R-quadrado | 0.720973 | Média dependente var. | | –0.047826 |
| R-quadrado ajustado | 0.607618 | S.D. dependente var. | | 0.484763 |
| S.E. regressão | 0.303658 | Akaike info. critério | | 0.699958 |
| Sum quadrados res. | 2.950653 | Schwarz critério | | 1.256501 |
| Log probabilidade | –2.099034 | Hannan-Quinn critério. | | 0.908442 |
| F-estatística | 6.360315 | Durbin-Watson stat. | | 2.270609 |
| Prob(F-estatística) | 0.000010 | | | |

**Fonte:** Os autores (resultados gerados pelo *Eviews*)

**Imagem 02** Correlograma dos resíduos do modelo 02

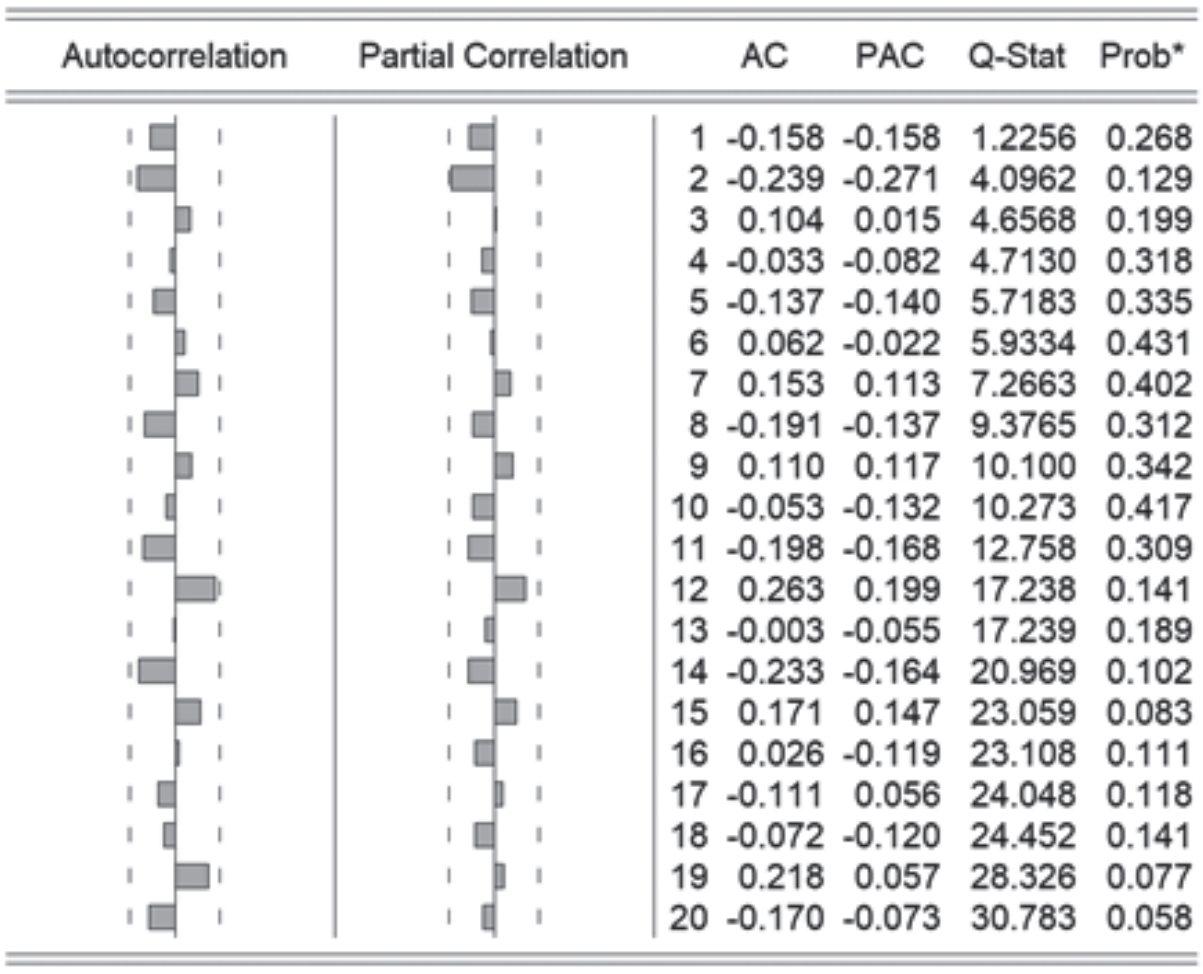

| Autocorrelation | Partial Correlation | | AC | PAC | Q-Stat | Prob* |
|---|---|---|---|---|---|---|
| | | 1 | -0.158 | -0.158 | 1.2256 | 0.268 |
| | | 2 | -0.239 | -0.271 | 4.0962 | 0.129 |
| | | 3 | 0.104 | 0.015 | 4.6568 | 0.199 |
| | | 4 | -0.033 | -0.082 | 4.7130 | 0.318 |
| | | 5 | -0.137 | -0.140 | 5.7183 | 0.335 |
| | | 6 | 0.062 | -0.022 | 5.9334 | 0.431 |
| | | 7 | 0.153 | 0.113 | 7.2663 | 0.402 |
| | | 8 | -0.191 | -0.137 | 9.3765 | 0.312 |
| | | 9 | 0.110 | 0.117 | 10.100 | 0.342 |
| | | 10 | -0.053 | -0.132 | 10.273 | 0.417 |
| | | 11 | -0.198 | -0.168 | 12.758 | 0.309 |
| | | 12 | 0.263 | 0.199 | 17.238 | 0.141 |
| | | 13 | -0.003 | -0.055 | 17.239 | 0.189 |
| | | 14 | -0.233 | -0.164 | 20.969 | 0.102 |
| | | 15 | 0.171 | 0.147 | 23.059 | 0.083 |
| | | 16 | 0.026 | -0.119 | 23.108 | 0.111 |
| | | 17 | -0.111 | 0.056 | 24.048 | 0.118 |
| | | 18 | -0.072 | -0.120 | 24.452 | 0.141 |
| | | 19 | 0.218 | 0.057 | 28.326 | 0.077 |
| | | 20 | -0.170 | -0.073 | 30.783 | 0.058 |

**Fonte:** Os autores (resultados gerados pelo *Eviews*)

No modelo 03 (p-valor a 05%), o *spread* é afetado positivamente pelos fatores; Inadimplência, PIB, taxa de desemprego e Selic. O modelo mostra que o *spread* é afetado negativamente pelo saldo total das operações de crédito com recursos livres, pois com um maior saldo, o volume de empréstimos é maior e, portanto, o lucro do banco também, de modo que os bancos podem, neste contexto, reduzirem as taxas de juros para atraírem mais tomadores de empréstimo. O $R^2$ da regressão foi de 76,90 %. Os resíduos do modelo mostram-se não autocorrelacionados. A tabela 06 e a imagem 03 a seguir, apresentam os detalhes estatísticos do modelo 03.

**Tabela 06** Resultados do modelo 03

| Variável | Coeficiente | Erro padr. | t-Estatística | Prob. |
|---|---|---|---|---|
| C | 0.162809 | 0.063258 | 2.573715 | 0.0141 |
| D(INADTOTAL) | 1.914049 | 0.695808 | 2.750828 | 0.0091 |
| D(CREDITOTOTAL) | –2.17E–05 | 5.23E–06 | –4.147609 | 0.0002 |
| D(PIB (-4)) | 8.67E–06 | 2.87E–06 | 3.019810 | 0.0045 |
| D(DESEMPREGO) | 0.335438 | 0.128810 | 2.604120 | 0.0131 |
| D(SELIC (-1)) | 0.422694 | 0.153843 | 2.747571 | 0.0091 |
| | | | | |
| R-quadrado | 0.769099 | Média dependente var. | | –0.045455 |
| R-quadrado ajustado | 0.738718 | S.D. dependente var. | | 0.495306 |
| S.E. regressão | 0.253179 | Akaike info. critério | | 0.216687 |
| Sum quadrados res. | 2.435793 | Schwarz critério | | 0.459985 |
| Log probabilidade | 1.232889 | Hannan-Quinn critério. | | 0.306914 |
| F-estatística | 25.31458 | Durbin-Watson stat. | | 1.854438 |
| Prob(F-estatística) | 0.000000 | | | |

**Fonte:** Os autores (resultados gerados pelo *Eviews*)

**Imagem 03** Correlograma dos resíduos do modelo 03

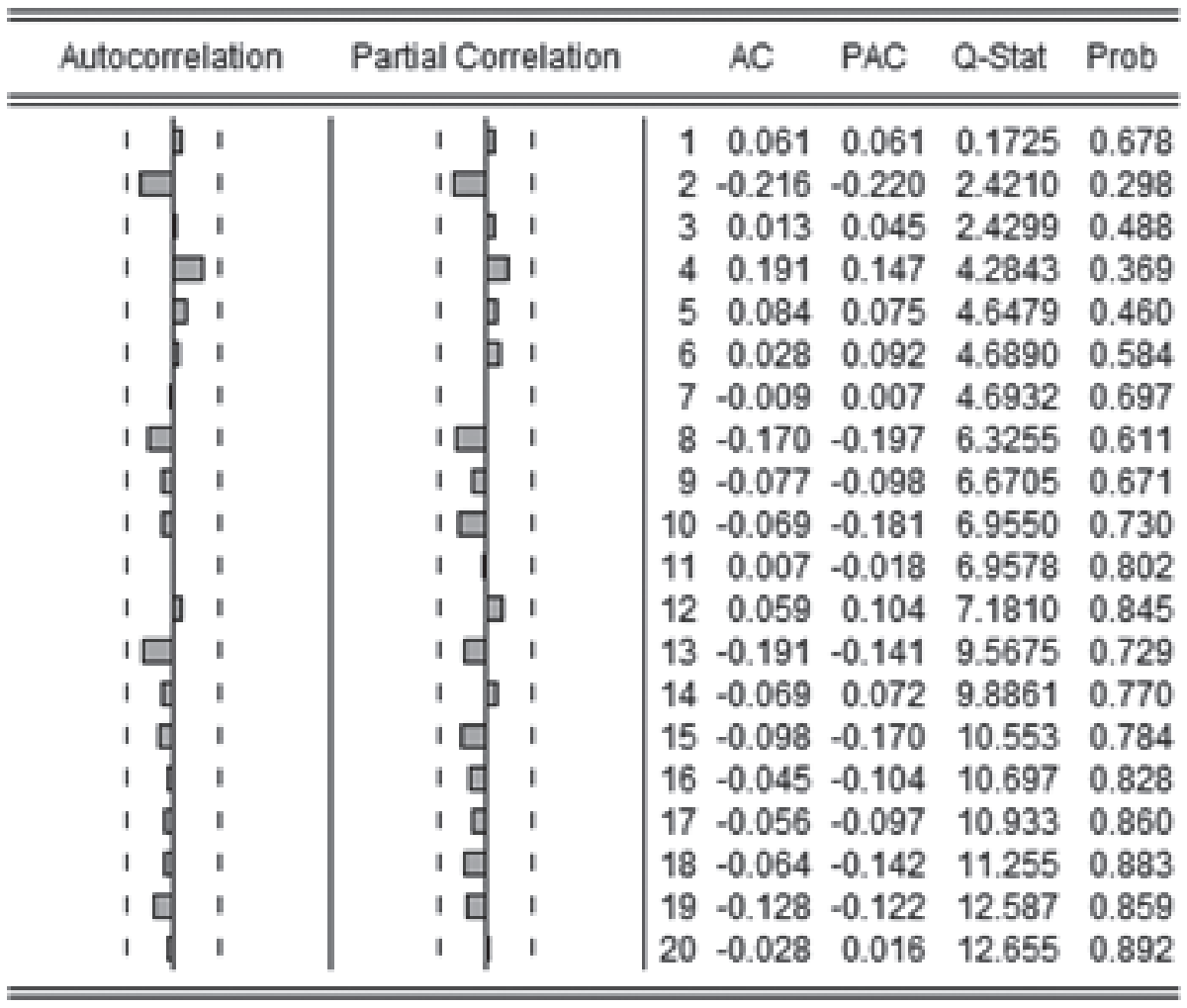

| Autocorrelation | Partial Correlation | | AC | PAC | Q-Stat | Prob |
|---|---|---|---|---|---|---|
| | | 1 | 0.061 | 0.061 | 0.1725 | 0.678 |
| | | 2 | -0.216 | -0.220 | 2.4210 | 0.298 |
| | | 3 | 0.013 | 0.045 | 2.4299 | 0.488 |
| | | 4 | 0.191 | 0.147 | 4.2843 | 0.369 |
| | | 5 | 0.084 | 0.075 | 4.6479 | 0.460 |
| | | 6 | 0.028 | 0.092 | 4.6890 | 0.584 |
| | | 7 | -0.009 | 0.007 | 4.6932 | 0.697 |
| | | 8 | -0.170 | -0.197 | 6.3255 | 0.611 |
| | | 9 | -0.077 | -0.098 | 6.6705 | 0.671 |
| | | 10 | -0.069 | -0.181 | 6.9550 | 0.730 |
| | | 11 | 0.007 | -0.018 | 6.9578 | 0.802 |
| | | 12 | 0.059 | 0.104 | 7.1810 | 0.845 |
| | | 13 | -0.191 | -0.141 | 9.5675 | 0.729 |
| | | 14 | -0.069 | 0.072 | 9.8861 | 0.770 |
| | | 15 | -0.098 | -0.170 | 10.553 | 0.784 |
| | | 16 | -0.045 | -0.104 | 10.697 | 0.828 |
| | | 17 | -0.056 | -0.097 | 10.933 | 0.860 |
| | | 18 | -0.064 | -0.142 | 11.255 | 0.883 |
| | | 19 | -0.128 | -0.122 | 12.587 | 0.859 |
| | | 20 | -0.028 | 0.016 | 12.655 | 0.892 |

**Fonte:** Os autores (resultados gerados pelo *Eviews*)

No modelo 04, a variável PIB foi excluída e em seu lugar foi adicionado o índice de volume de vendas no varejo, com o propósito de analisar possível influência do consumo sobre o *spread*. Considerando-se p-valor a 05% praticamente todas as variáveis são significativas, somente o desemprego que entraria a um p-valor de 10%. O *spread* total é afetado positivamente; pela inadimplência, pela taxa básica de juros, a Selic, que serve como base para as taxas de juros praticadas pelos bancos. No modelo 04 o *spread* é afetado negativamente; pelo saldo total das operações de crédito com recursos livres, e pelo nível de atividade econômica, medido por meio do índice de volume de vendas no varejo com cinco defasagens. O $R^2$ da regressão foi de 76,85 %. Os resíduos deste modelo também se mostram não autocorrelacionados. A tabela 07 e a imagem 04 a seguir, apresentam os detalhes estatísticos do modelo 04.

**Tabela 07** Resultados do modelo 04

| Variável | Coeficiente | Erro padr. | t-Estatística | Prob. |
|---|---|---|---|---|
| C | 0.175099 | 0.063502 | 2.757377 | 0.0090 |
| D(INADTOTAL) | 3.156704 | 0.837587 | 3.768806 | 0.0006 |
| D(CREDITOTOTAL) | –1.79E–05 | 5.65E–06 | –3.159466 | 0.0031 |
| D(DESEMPREGO) | 0.230095 | 0.129221 | 1.780623 | 0.0832 |
| D(SELIC (-1)) | 0.621436 | 0.168031 | 3.698349 | 0.0007 |
| D(VENDA_VAREJO(-5)) | –0.009458 | 0.003210 | –2.946179 | 0.0055 |
| | | | | |
| R-quadrado | 0.768490 | Média dependente var. | | –0.044186 |
| R-quadrado ajustado | 0.737205 | S.D. dependente var. | | 0.501095 |
| S.E. regressão | 0.256879 | Akaike info. critério | | 0.248365 |
| Sum quadrados res. | 2.441514 | Schwarz critério | | 0.494114 |
| Log probabilidade | 0.660150 | Hannan-Quinn critério. | | 0.338990 |
| F-estatística | 24.56408 | Durbin-Watson stat. | | 1.766322 |
| Prob(F-estatística) | 0.000000 | | | |

**Fonte:** Os autores (resultados gerados pelo *Eviews*)

**Imagem 04** Correlograma dos resíduos do modelo 04

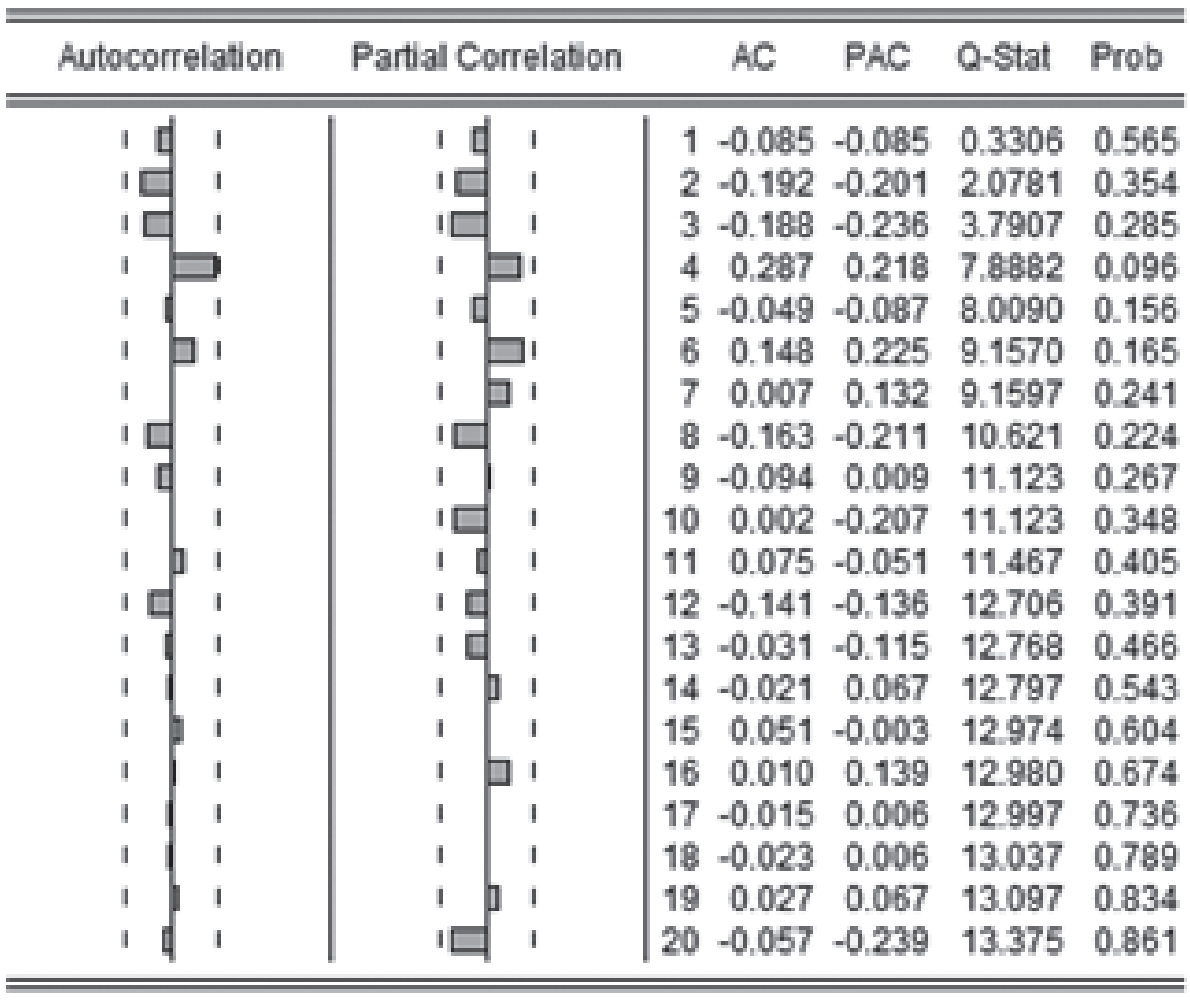

| Autocorrelation | Partial Correlation | | AC | PAC | Q-Stat | Prob |
|---|---|---|---|---|---|---|
| | | 1 | -0.085 | -0.085 | 0.3306 | 0.565 |
| | | 2 | -0.192 | -0.201 | 2.0781 | 0.354 |
| | | 3 | -0.188 | -0.236 | 3.7907 | 0.285 |
| | | 4 | 0.287 | 0.218 | 7.8882 | 0.096 |
| | | 5 | -0.049 | -0.087 | 8.0090 | 0.156 |
| | | 6 | 0.148 | 0.225 | 9.1570 | 0.165 |
| | | 7 | 0.007 | 0.132 | 9.1597 | 0.241 |
| | | 8 | -0.163 | -0.211 | 10.621 | 0.224 |
| | | 9 | -0.094 | 0.009 | 11.123 | 0.267 |
| | | 10 | 0.002 | -0.207 | 11.123 | 0.348 |
| | | 11 | 0.075 | -0.051 | 11.467 | 0.405 |
| | | 12 | -0.141 | -0.136 | 12.706 | 0.391 |
| | | 13 | -0.031 | -0.115 | 12.768 | 0.466 |
| | | 14 | -0.021 | 0.067 | 12.797 | 0.543 |
| | | 15 | 0.051 | -0.003 | 12.974 | 0.604 |
| | | 16 | 0.010 | 0.139 | 12.980 | 0.674 |
| | | 17 | -0.015 | 0.006 | 12.997 | 0.736 |
| | | 18 | -0.023 | 0.006 | 13.037 | 0.789 |
| | | 19 | 0.027 | 0.067 | 13.097 | 0.834 |
| | | 20 | -0.057 | -0.239 | 13.375 | 0.861 |

**Fonte:** Os autores (resultados gerados pelo *Eviews*)

A tabela 08 resume as principais observações dos modelos 01, 02, 03 e 04.

A tabela 09 a seguir, mostra as variáveis que estatisticamente são capazes de reduzir o *spread*, ou seja, há influência negativa. Conforme há aumento da expressão dessas variáveis, tende a ser observada redução do *spread* (p-valor a 05%).

**Tabela 09** Variável independente com padrão negativo de influência

| Variáveis. | Modelos |
|---|---|
| IPI bens de consumo | 01 e 02 |
| IPI geral | 02 |
| IPCA (t-3) | 02 |
| Saldo total das op. de crédito (recursos livres) | 03 e 04 |
| Índice do volume de vendas no varejo | 04 |

**Fonte:** elaborada pelos autores

**Tabela 08** Resumo dos resultados dos modelos

| Modelo | R-Quadrado | Autocorrelação dos resíduos | Parâmetros significantes a 05% |
|---|---|---|---|
| 01 | 68,19% | Inexistente | EMBI+, Inadtotal, Ipi_benscapital, ipi_bensconsumo, ipi_bensconsumodur, ipibensintermediarios, ipi_bensnaoesemidur. |
| 02 | 72,09 % | Inexistente | EMBI+, Inadtotal, IPCA (t-3), Ipi_benscapital, ipi_bensconsumo, ipi_bensconsumodur, ipibensintermediarios, ipi_bensnaoesemidur e ipi_geral e Selic. |
| 03 | 76,90 % | Inexistente | Inadtotal, PIB, taxa de desemprego, Selic, e saldo de crédito das operações de recursos livres. |
| 04 | 76,85 % | Inexistente | Inadtotal, Selic, saldo de crédito das operações de recursos livres, e índice do volume de vendas no varejo. |

**Fonte:** Elaborada pelos autores

## 5 DISCUSSÃO

No presente estudo, dentre as variáveis candidatas a influenciar o *spread* no período analisado, quatro não se mostraram estatisticamente significantes, são elas; IGPM, índice de produção de máquinas agrícolas, indicador de produção da indústria extrativa-mineral e o indicador de produção da indústria da transformação. Todas as demais variáveis empregadas, para os quatro modelos, apresentaram evidências estatísticas de significância sobre o *spread*, com p-valor a 05%. São significantes; IPI geral (no modelo 02 quando IPCA é defasado em três períodos), inadimplência total, IPI para bens de capital, IPI bens intermediários, IPI bens de consumo, IPI bens de consumo duráveis, IPI bens semiduráveis e não duráveis, o IPCA (modelo 02), Selic, PIB, saldo da carteira de crédito total, índice de volume de vendas no varejo total, o desemprego (modelo 03) e EMBI+, com p-valor a 05%.

Os setores industriais significantes (p-valor a 05%) apresentam evidências estatísticas de influen-ciar o *spread* positivamente. Entretanto, o IPI bens de consumo (modelos 01 e 02) teve o mesmo comportamento do volume de vendas no varejo (modelo 04) e das operações e crédito (modelos 03 e 04). O indicador geral (IPI geral no modelo 02) também apresenta influencia negativa sobre o *spread* bancário, este padrão de influência para o IPI geral também fora observado por Matulovic (2015). Assim, em um contexto com elevado nível da produção industrial geral, , estabilidade econômica, aumento da oferta de crédito e baixa inadimplência, pode ser observada a redução do *spread*  A relação de influência negativa do IPI bens de consumo com o *spread* também é coerente em momentos de prospecção de consumo, maior vendas no varejo e menor inadimplência.

Em relação ao modelo 01 é coerente em um contexto econômico existir um maior *spread* bancário quando os indicadores EMBI+, inadimplência, desemprego e Selic estiverem elevados, pois existem evidências de maiores probabilidades de riscos e vulnerabilidades macroeconômicas que os bancos podem vivenciar neste cenário. Quanto aos indicadores industriais que mostram influenciar positivamente o *spread*, pode ser esperado um aumento da demanda por crédito, e consequente elevação do *spread*, em função da expansão dessas atividades industriais (IPI bens de capital, IPI bens de consumo duráveis, IPI bens intermediários e IPI bens não e semiduráveis) o que é coerente com as evidências estatísticas encontradas neste estudo.

O modelo 04 pode ser mais interessante do que o modelo 03 por considerar o índice do volume de vendas no varejo.. O crédito total e o índice do volume de vendas no varejo possuem relação de influência negativa sobre o *spread* bancário. É economicamente coerente afirmar que quanto maior o volume de crédito total, menor tende a ser o *spread* bancário, pois nesta situação há evidências de um ambiente macroeconômico mais estável e com baixa inadimplência. Neste contexto, o consumo geral tende a ser maior, sendo também coerente o comportamento estatístico do volume de vendas no varejo em relação ao padrão de influência (negativo) dos indicadores de produção industrial geral (modelo 02) e de consumo (modelos 01 e 02).

Há evidências para argumentar que quanto maior for o volume de vendas no varejo, maior tende a ser a produção industrial geral e também o nível da produção de bens de consumo. Em contexto de baixo desemprego e menor inadimplência, é economicamente coerente esperar existir maior produção industrial, maior consumo, maior crédito liberado na economia, acompanhados do progresso das vendas no varejo, o que segundo as evidências estatísticas encontradas neste estudo, permitem afirmar que haverá um menor *spread* bancário. Em relação ao PIB, no presente estudo foi observada relação positiva e significativa de influência sobre o *spread*, esta observação também fora constatada por Almeida e Divino (2013). O PIB elevado tende a influenciar os consumidores na busca por crédito, ocorrendo aumento da demanda, o que será incorporado nos modelos de precificação bancários.

Os padrões de influência das variáveis utilizadas no presente estudo, sobre o *spread* bancário no Brasil, reafirmam observações previamente pautadas na literatura, a exemplo; Selic (+) por Koyama e Nakane (2001a e 2002b), Bignotto e Rodrigues (2006), Oreiro et al (2006). EMBI+ (+) por Auel e Mendonça (2011), Pereira Tavares *et al* (2013). Desemprego (+) por Manhiça e Jorge (2012). IPI geral (-) por Koyama e Nakane (2001), Matulovic (2015). PIB (+) por Koyama e Nakane (2001), Almeida e Divino (2013). IPCA (-) por Bignotto e Rodrigues (2006). Neste trabalho, o comportamento estatístico dos indicadores de produção industrial (IPIs) e da inadimplência, significativos a p-valor de 05%, permitem confirmar a constatação de Souza e Coelho (2008) de que a atividade industrial influencia o *spread* bancário no Brasil.

## 6 CONCLUSÃO E IMPLICAÇÕES

Neste estudo os fatores que influenciam o *spread* são; o IPI geral (-), o IPCA (-), a inadimplência total (+), o IPI para bens de capital (+), IPI bens intermediários (+), IPI bens de consumo (-), IPI bens de consumo duráveis (+), IPI bens semiduráveis e não duráveis (+), a Selic (+), o PIB (+), o saldo da carteira de crédito total (-), o índice de volume de vendas no varejo (-), a taxa de desemprego região metropolitana (+) e o EMBI+ (+), considerando p-valor a 05%. Foi possível constatar que os indicadores de produção industrial (com significância estatística) exercem influência sobre o *spread.*

Os fatores que influenciam o *spread* de modo negativo são; o IPI para bens de consumo (modelos 01 e 02) e IPI geral (modelo 02), o IPCA (t-3) (modelo 02), o saldo da carteira de crédito total para recursos livres (modelos 03 e 04) e índice de volume de vendas no varejo (modelo 04). Os demais fatores encontrados como estatisticamente determinantes do *spread* bancário o influenciam positivamente. O presente estudo contribui para evidenciar que há relação estatística significativa de alguns dos indicadores industriais sobre o *spread*. Este trabalho também mostra a influência por parte dos indicadores macroeconômicos, sendo eles; IPCA (-), inadimplência (+), Selic (+), PIB (+), saldo da carteira de crédito total (-), volume de vendas no varejo (-), desemprego (+) e EMBI+ (+).

As maiores contribuições desta pesquisa para a literatura e para a sociedade podem ser resumidas em duas principais constatações. Há evidências estatísticas de que é possível esperar um menor *spread* e uma melhor eficiência na alocação de recursos na economia quanto mais prósperos estiverem; o nível de atividade industrial geral, o volume de vendas no varejo, o volume de crédito disponível (recursos livres) e o nível da indústria de bens de consumo. A segunda principal observação é que há relação de influência da atividade industrial, em diferentes subclassificações, sobre o *spread* bancário. Assim, considerando-se que o *spread* pode ser adotado como um indicador de eficiência no processo de intermediação financeira, as constatações do presente estudo colaboram como base para o desenvolvimento de futuras estratégias econômicas as quais estimulem o progresso das indústrias e da competitividade das empresas brasileiras.

A principal limitação durante a construção desta pesquisa consiste na escassez de trabalhos acadêmicos e científicos que abordem o tema de *spread* bancário (relacionado à indústria nacional) na literatura brasileira. Esta pesquisa contribui para a literatura ao abordar diretamente os indicadores industriais e suas respectivas subclassificações, com o *spread* bancário brasileiro.

Em pesquisas futuras, outras variáveis poderiam ser testadas, como; a participação pública e a privada sobre o crédito total, a Selic real, em paralelo à Selic nominal, indicadores de tecnologia (ou indústria da tecnologia); além de considerar, nas modelagens, diferenciação entre *spreads* de bancos públicos e privados. Estudar mais detalhadamente a influência por segmento industrial sobre o *spread*, também agregará a literatura.

Embora os setores relacionados ao agronegócio e a extração de minério de ferro representem importantes commodities para a economia, nesta pesquisa há evidências de que estes segmentos não influenciam o *spread* bancário, conforme comportamento estatístico dos IPIs da indústria de máquinas agrícolas e extrativa-mineral. Entretanto, em estudos futuros podem ser abordados especificamente fatores microeconômicos e intrínsecos a estes setores e suas relações com o *spread*. O agronegócio, por exemplo, possui muitas volatilidades econômicas devido aos ciclos de safras e desempenhos de produção. Estas oscilações são certamente incorporadas aos componentes de risco nas modelagens de precificação bancárias.

A presente pesquisa permite afirmar que o progresso da indústria brasileira, o aumento das vendas no varejo e concomitante prospecção do consumo nacional em contexto de geração de empregos e estabilidade econômica, são fatores de sucesso para ocorrer à redução do *spread* bancário e melhora na eficiência no processo de intermediação financeira. Neste cenário, as empresas e indústrias brasileiras encontram melhores forças para o ganho de competitividade.

## REFERÊNCIAS

AFANASIEFF, T. S.; LHACER, P. M.; NAKANE, M. I. The determinants of bank interest spread in Brazil. Money Affairs, v. 15, n.2, p. 183-207, 2002.

ARONOVICH, S. Uma nota sobre os efeitos da inflação e do nível de atividade sobre o *spread* bancário. Revista Brasileira de Economia, v. 48, n. 1, p.125-140, 1994.

AUEL, M. C.; MENDONÇA, H. F. Macroeconomic relevance of credit channels: Evidence from an emerging economy under inflation targeting. Economic Modelling, p. 965–979, 2011.

Banco Central do Brasil (BACEN). Informações obtidas por acesso ao endereço eletrônico do Banco Central do Brasil. www.bcb.gov.br. 2015.

BIGNOTTO, F.; RODRIGUES, E. Fatores de risco e spread bancário no Brasil. Trabalhos para Discussão do Banco Central do Brasil, n. 110.2006.

CHORTAREAS, G. E., GARZA-GARCIA, J. G.;GIRARDONE, C. Competition, efficiency and interest rate margins in Latin American banking. International Review of Financial Analysis, v.24, p. 93-103, 2012.

DANTAS ALMEIDA, F.; DIVINO, J. A. Determinantes do *spread* bancário *ex-post* no Brasil: Uma análise de fatores micro e macroeconômicos. Dissertação de mestrado em economia. Universidade Católica de Brasília, 2013.

DANTAS, J.A., MEDEIROS, O.R.; CAPELLETTO, L.R. Determinantes do *spread* bancário *ex-post* no mercado brasileiro. Trabalhos para discussão do Banco Central do Brasil, 2011.

DEMIRGUÇ-KUNT, A.; HUIZINGA, H. Determinants of commercial bank interest margins and profitability: Some international evidence. World Bank Economic Review, v. 13, n. 2, p. 379-408, 1999.

HO, T. S. Y.; SAUNDERS, A. The determinants of bank interest margins: Theory and empirical evidence. Journal of Financial and Quantitative Analysis, v. 16, n.4, p. 581-600, 1981.

JORGENSEN, O. H.; APOSTOLOU, A. Brazil's bank spread in international context from micro to macro level. Policy Research Working Paper 661. The World Bank Latin America and the Caribbean region, poverty reduction and economic management department, 2013.

KLEIN, M. A. A theory of the banking firm. Journal of Money, Credit and Banking, v.2, n. 3, p. 205-218, 1971.

KOYAMA, S. M.; NAKANE, M. I. (2001a). Os determinantes do Spread bancário no Brasil. In: Banco Central do Brasil, juros e Spread bancário no Brasil: avaliação de dois anos do projeto. Brasília: Banco Central do Brasil, pp.27-30.2001.

KOYAMA, S. M.; NAKANE, M. I. O spread bancário segundo fatores de persistência e conjuntura. Relatório de Economia Bancária e Crédito, Novembro de 2001. Banco Central do Brasil. 2001.

KOYAMA, S.; NAKANE, M. (2002a). O spread bancário segundo fatores de persistência e conjuntura. *Notas Técnicas do Banco Central do Brasil*, n. 18.2002.

KOYAMA, S.; NAKANE, M. (2002b). Os determinantes do spread bancário no Brasil. Notas Técnicas do Banco Central do Brasil, n. 19. 2002.

LEAL DE SOUZA, R. M. Estrutura e determinantes do *spread* bancário no Brasil: Uma resenha comparativa da literatura empírica. Dissertação de Mestrado. Faculdade de Ciências Econômicas – Universidade do Estado do Rio de Janeiro, 2006.

MANHIÇA, F. A.; JORGE, C. T. O nível da taxa básica de juros e o *spread* bancário no Brasil: Uma análise de dados em painel. Texto para Discussão 1710, IPEA, 2012.

MATULOVIC, M. O. Os Determinantes Macroeconômicos do *spread* bancário para pessoas físicas e jurídicas no Brasil: Uma análise do período pós-Plano Real. Dissertação de Mestrado. Escola de Economia de São Paulo – Fundação Getúlio Vargas, 2015.

MUELLER, A. Uma aplicação de redes neurais artificiais na previsão do mercado acionário. Dissertação (Mestrado em Engenharia de Produção) – Programa de Pós-graduação em Engenharia de Produção. Universidade Federal de Santa Catarina (UFSC), 1996.

NAKANE, M. A test of Competition in Brazilian Banking. Working Paper Series. Central Bank of Brasil. 2001.

NAKANE, M.; COSTA, A. *Spread* bancário: Os problemas da comparação internacional. Risk Update, v.1, n.3, p. 9-14, 2005.

OREIRO, J.L.; DE PAULA, L.F.R; COSTA DA SILVA, G.J.; AMARAL, R.Q. Por que o Custo do Capital no Brasil é Tão Alto? Revista de Economia Política, vol. 32, nº 4 (129), pp. 557-579, outubro-dezembro/2012.

OREIRO, J.; PAULA, L.; ONO, F.; COSTA DA SILVA, G. Determinantes macroeconômicos do *spread* bancário no Brasil: Teoria e evidência recente. Economia Aplicada, v. 4, n. 10, p. 609-634, 2006.

PAIM, B.O Sistema Financeiro Nacional de 2008 a 2013: A Importância das Instituições Públicas. Núcleo de Estudos e Política Econômica (NEPE) FEE, v.2, n.41, p. 25-40, 2013.

PEREIRA TAVARES, D., CALDAS MONTES, G.; DE CARVALHO GUILLÉN, O. T. Transmissão da política monetária pelos canais de tomada de risco e de crédito: Uma análise considerando os seguros contratados pelos bancos e o *spread* de crédito no Brasil. Revista Brasileira de Economia (RBE), v.3, n. 67, p. 337-353, 2013.

SOUZA, F.S.M.; COELHO, J. C. Determinantes do *spread* bancário e da concessão de crédito no Brasil. Universidade Federal de Santa Catarina, 2008.

WORLD BANK; IMF. Indicators of financial structure, development, and soundness. In: Financial sector Assessment: a handbook, v.2, p. 15-33, 2005.